\RequirePackage{fix-cm}
\documentclass[smallextended]{svjour3}       
\smartqed  
\usepackage{graphicx}
\usepackage{amsmath}
\usepackage{amssymb}

\newcommand{\beq}{\begin{equation}}
\newcommand{\eeq}{\end{equation}}
\newcommand{\bey}{\begin{eqnarray}}
\newcommand{\eey}{\end{eqnarray}}
\newcommand{\bal}{\begin{align}}
\newcommand{\eal}{\end{align}}

\newcommand{\R}{\mathbb{R}}

\newcommand{\PP}{\mathbf{P}}
\newcommand{\pp}{\mathbf{p}}

\newcommand{\XX}{\mathbf{X}}
\newcommand{\GG}{\mathbf{G}}
\newcommand{\LL}{\mathbf{L}}
\newcommand{\bpi}{\mathbf{\Pi}}

\newcommand{\CH}{\mathcal{H}}

\newcommand{\bra}[1]{\langle #1 }
\newcommand{\ket}[1]{, #1 \rangle}
\newcommand{\bs}[1]{\boldsymbol{#1}}
\newcommand{\Hilb}{\mathcal{H}}

\begin{document}

\title{Quantum theory of the Generalised Uncertainty Principle
}
\titlerunning{GUP and minimal length}

\author{Jean-Philippe Bruneton        \and
        Julien Larena 
}


\institute{Jean-Philippe Bruneton \at
              LUTH, Observatoire de Paris, PSL Research University, CNRS, Université Paris Diderot, Sorbonne Paris Cit\'e, 5 place Jules Janssen, 92195 Meudon Paris, France\\
              \email{jean-philippe.bruneton@obspm.fr}           
           \and
           Julien Larena \at
              Department of Mathematics and Applied Mathematics, University of Cape Town, Rondebosch 7701, South Africa\\ 
              \email{julien.larena@uct.ac.za}
}

\date{Received: date / Accepted: date}

\maketitle

\begin{abstract}
We extend significantly previous works on the Hilbert space representations of the Generalized Uncertainty Principle (GUP) in 3+1 dimensions of the form $[X_i,P_j] = i F_{ij}$ where $F_{ij} = f(\PP^2) \delta_{ij} + g(\PP^2) P_i P_j$ for any functions $f$. However, we restrict our study to the case of commuting $X$'s. We focus in particular on the symmetries of the theory, and the minimal length that emerge in some cases.
We first show that, at the algebraic level, there exists an unambiguous mapping between the GUP with a deformed quantum algebra and a quadratic Hamiltonian into a standard, Heisenberg algebra of operators and an aquadratic Hamiltonian, provided the boost sector of the symmetries is modified accordingly. The theory can also be mapped to a completely standard Quantum Mechanics with standard symmetries, but with momentum dependent position operators. Next, we investigate the Hilbert space representations of
these algebraically equivalent models, and focus specifically on whether they exhibit a minimal length. We carry the functional analysis of the various operators involved, and show that the
appearance of a minimal length critically depends on the relationship between the generators of translations and the physical momenta. In particular, because this relationship is preserved by the algebraic mapping presented in this paper, when a minimal length is present in the standard GUP, it is also present in the corresponding Aquadratic Hamiltonian formulation, despite the perfectly standard algebra of this model.
In general, a minimal length requires bounded generators of translations, i.e. a specific kind of quantization of space, and this depends on the precise shape of the function $f$ defined previously. This result provides an elegant and unambiguous classification of which universal quantum gravity corrections lead to the emergence of a minimal length.
\keywords{Generalised Uncertainty Principle \and Minimal Length scenarios}

\PACS{03.65.-w, 03.65.Db, 04.60.Bc}

\end{abstract}

\section{Introduction}
The existence of a minimal length of the order of the Planck scale has attracted a lot of attention recently. Its purported existence comes mainly from thought experiments that combine both gravitational and quantum ingredients \cite{mead1964,Maggiore:1993rv,Adler:1999bu,Scardigli:1999jh,Padmanabhan:1987au,Garay:1994en,hossenfelder2013minimal,burderi:2012ze,Bruneton:2013ena}, but similar results have been claimed to derive from candidates to a theory of quantum gravity; e.g. string theory \cite{Konishi:1989wk,Amati:1988tn}. Even earlier, Snyder \cite{snyder:1946qz} provided some arguments for alterations to quantum mechanics in the quantum gravity regime.

A minimal length can be studied in a somewhat purely classical context, in which case it has to be another invariant of the symmetries and it gives rise to models such as Deformed Special Relativity and/or relative locality (see, e.g. \cite{hossenfelder:2005ed,Girelli:2007sz,Ghosh:2007ai,AmelinoCamelia:2011bm,AmelinoCamelia:2011pe}). Due to its quantum mechanical origin though, it is more natural to study it at the quantum level directly. This has led to the so called Generalized Uncertainty Principle (GUP), where an extra term in the right hand side of Heisenberg inequality forces the existence of a minimal variance for the measurement of distances. The generalized commutation relation between position and momenta from which it derives has attracted a lot of attention in the last decade.

Besides the question of its Hilbert space representations, first studied in the famous paper \cite{Kempf:1994su}, many following studies have focussed on potential phenomenological implications, e.g. modifications to Newton's law \cite{Ali:2013ma}, deformed measure on the phase space and its thermodynamical consequence \cite{Chang:2001bm,Nozari:2006au,Ali:2014dfa}, atomic physics (harmonic oscillator, hydrogen atom) \cite{Kempf:1996fz,das:2009hs,Pedram:2010hx,Ali:2011fa,Ghosh:2011ze}, relation with the Weak Equivalence Principle \cite{ali:2011ap,Ghosh:2013qra,Das:2014bba}, cosmology and early cosmology \cite{nozari:2006gg,battisti:2007jd,Vakili:2008tt,Tawfik:2012he}, Black Hole physics \cite{medved:2004yu,adler:2001vs,bina:2010ir,myung:2006qr,Zhao:2006xf,kim:2007hf,Xiang:2009yq,Yang:2009vf,Majumder:2011xg,Scardigli:2014qka,Faizal:2014tea,Ali:2015zua}, relation to holography \cite{scardigli:2003kr,Kim:2008kc}, or even to the notion of time \cite{Faizal:2014mba}. Note also a recent, original proposal to link the GUP to the putative existence of sub-Planckian Black Holes \cite{Carr:2015nqa}. For more comprehensive reviews, in particular with respect to applications of the GUP to physical problems, see \cite{Tawfik:2014zca,Tawfik:2015rva}.

Finally, extensions of the GUP to a modified Quantum Field Theory have also been explored \cite{Matsuo:2005fb,Husain:2012im,Kempf:1996nk}. All this phenomenology relies on the fact that the GUP generically leads to a discretisation of space \cite{ali:2009zq,das:2010zf,Balasubramanian:2014pba,Deb:2016psq}, and that the converse is also true \cite{bang:2006va}.

These phenomenological studies usually consider a standard, quadratic Hamiltonian, $H=P^2/2m$, plus some potential in 1+1 dimension, where $P$, identified with the physical momentum, is not equal to the translation operator $\Pi$ any more, leading to a Generalized Commutation Relation (GCR) of the form: 
\begin{equation}
\label{GUPIntro}
[X,P]= i f(P)
\end{equation}
between position and momentum operators. It is natural to ask why a quadratic Hamiltonian should be a preferred choice in this case. Indeed, as it is well-known, the Hamiltonian determines (or is determined by, depending on the point of view) the symmetries of the model, and there must then exists a link between the functional form of $H$, the GCR, and the symmetries.

In this paper we show explicitly that such a relation exists. There is in fact an infinite number of ways to write (at least at the algebraic level), the GUP theory of the free particle in 3+1 dimensions, and its symmetries. This ranges from a theory which is basically Quantum Mechanics with a non quadratic Hamiltonian (thereafter referred to as Aquadratic QM, or AQM), to its equivalent GUP (algebraic) representation written above, and even to standard Quantum Mechanics, with quadratic Hamiltonian, standard Galilean symmetries, but acting on modified, momentum dependent space operators $\tilde X = F(X,P)$, reminiscent of Deformed Special Relativity (DSR) models.

This construction is straightforward and rather trivial from the purely algebraic point of view, i.e. as long as one only considers the algebras of operators including the position operator, the momentum operator, and the generators of symmetries (translation, rotations and Galilean boosts). However, when considered as different sets of linear operators acting on different Hilbert spaces, i.e. when one tries to represent these algebras on Hilbert spaces characterising physical states, their equivalence/non-equivalence is non trivial, and difficult to explicitly assess in general, as it requires a thorough study of a well-defined functional calculus on Hilbert spaces. We shall try and lay the basis for such a discussion in the end of the paper, leaving open questions for future work.

In Section 2, we recall the exact definition of GUP theories of the free particle in 3+1 dimensions, and we also provide a generalization of GCRs that encompasses all the modifications found so far in the literature (to the best of our knowledge). We then recall how symmetries are to be built from the Hamiltonian and give the explicit deformation of Galilean symmetries as a function of the GCRs.

In Section 3, we then derive explicitly the infinite family of equivalent theories at the algebraic level, emphasising in particular the AQM and the DSR like theories, and paying attention to the properties and the action of boosts. 

Finally, Section 4 deals with the question of representations on the Hilbert spaces. We first extend the work of \cite{Kempf:1994su} of the GUP representation in momentum space in 3+1 dimensions. We discuss the self-adjointness of the position operators and show how it depends critically on the boundedness or unboundedness of the range (or spectrum) of the translation operators, and therefore on functional properties of the GUP function $f$. We further show how this impacts the existence or not of a Minimal Length. We then construct the Hilbert space representation of the AQM, and discuss whether such properties transpose to this case. The DSR like model is however left for future works. 

Phenomenological implications will thus critically depend on the aforementioned properties of the function $f$. However, a detailed analysis of simple physical system (e.g. the particle in a box) with respect to these properties are left for a separate publication.

Capital letters are used to denote operators, while lower case letters denote c-numbers. Bold faced symbols denote vectors. Throughout the paper, $\bra{\psi}\ket{\phi}$ will denote the appriopriate scalar product on the Hilbert space $\mathcal{H}$ between $(\phi,\psi)\in\mathcal{H}^{2}$.

\section{3+1 GUP and its symmetries}

\subsection{General Algebra}
We start with a yet unspecified Hilbert space $\Hilb$ and an algebra of operators $\XX=(X_{1},X_{2},X_{3})$ and $\PP=(P_{1},P_{2},P_{3})$ on this Hilbert space, representing (in a yet unspecified manner) respectively a 'position' and a 'momentum' operators for a particle living in $\R^{3}$.
All cases in the literature so far are covered by the following restricted form of the algebra (see e.g. \cite{Maggiore:1993rv,Maggiore:1993kv,Pedram:2012my}):
\begin{eqnarray}
\label{Alg1}
[X_i,P_j]&\equiv& i F_{ij} = \left(f(\PP^2)\delta_{ij} + g(\PP^2) P_i P_{j}\right) \\
\label{Alg2}
[P_{i},P_{j}]&=& 0\\
\label{Alg3}
[X_{i},X_{j}]&=& i Z(\PP^2) \left( X_i P_j - X_j P_i\right)
\end{eqnarray}
where $\PP^{2}$ stands for $P_{1}^{2}+P_{2}^{2}+P_{3}^{2}$, and $g$, $Z$ and $f$ are analytic functions of their argument. We also work in natural units where the Planckian momentum is one. Note that the Hilbert space $\Hilb$ is unspecified because, as will be seen later, its structure depends strongly on the representation of the above algebra of operators on it. The operators of such an algebra satisfy the Jacobi identities if and only if:
\beq
\label{JacId}
Zf=-2ff'-2xf'g+fg,
\eeq
where the prime denotes a derivative of the functions with respect to their argument. In this paper, we further restrict the algebra to what we shall call the ``commutative GUP'', with $[X_{i},X_{j}]=0$, i.e. $Z\equiv 0$. Therefore, the relation (\ref{JacId}) fixes one functional degree of freedom, say $g$:
\beq
\label{Relfg}
g=\frac{2ff'}{f-2xf'} \mbox{ if } f-2xf'\neq 0.
\eeq
In the case where $f-2xf' = 0$, the fact that $Z=0$ in fact imposes $f=0$, leaving only $g \neq 0$ and arbitrary. We discard this case in the following as it cannot reduce to the standard QM in any limit.

\subsection{Hamiltonian and symmetries}

We now want to enrich the theory by supplementing the above algebra with a Hamiltonian $H(\XX,\PP)$ describing the dynamics of a particle. The symmetries of the model (defined on the Hilbert space $\Hilb$) will be generated by unitary operators $U$ which leave the Schr\"odinger equation invariant. This leads to:
\beq
i \frac{\partial U}{\partial t} = [H,U].
\eeq
Introducing the (Hermitian) generators of the transformations, $T_{i}$ such that $U \equiv e^{i \alpha_i T_i}$,
this leads to:
\beq
i \frac{\partial T_i}{\partial t} = [H,T_i].
\eeq

The notion of space carried by the operators $X_i$ enable us to define a notion of (invariance by) translation and rotations via the correspondence principle. For any state vector $\psi\in\CH$ and its image under the transformation $U$, $\psi'=U\psi$, we define:
\begin{itemize}
\item Translations in space:
\beq
\label{translation}
\bra{\psi'}\ket{\bs{X}\psi'}=\bra{\psi}\ket{\bs{X}\psi}+\bs{a},
\eeq
where $\bs{a}\in\mathbb{R}^{3}$ is the translation vector in ``classical'' space.
\item Rotations in space:
\bey
\bra{\psi'}\ket{\bs{X}\psi'}&=&\bs{R}\bra{\psi}\ket{\bs{X}\psi}\\
\bra{\psi'}\ket{\bs{P}\psi'}&=&\bs{R}\bra{\psi}\ket{\bs{P}\psi},
\eey
where $\bs{R}\in O(3)$ is the orthogonal matrix (e.g. in terms of Euler angles) representing the rotation in ``classical'' space. 
\end{itemize}

In each case, the unitary operators can be written $U_{T}=e^{ia^{j}\Pi_{j}}$ for translations and $U_{R}=e^{i\alpha^{j} L_{j}}$ for rotations. Working at first order, we can then determine the Hermitian generators of translation and rotations, $\Pi_{i}$ and $L_{i}$, which then obey:
\bey
&[X_{i},\Pi_{j}]&=i\delta_{ij}\\
&L_{i}&=\sum_{j,k}\epsilon_{ijk} X_{j}\Pi_{k},
\eey
where $\epsilon_{ijk}$ is the totally anti-symmetric tensor with $\epsilon_{123}=1$. In that case, it is easy to see, using the algebra for $\XX$ and $\PP$, that the generators of translation is related to the momentum of the particle through\footnote{A word of caution: this is valid in 3+1. In one-dimensional GUP of the form $[X,P]= i F_{1+1}(P)$, quite often considered in the literature, the generator of translations is rather given by 
$$
\Pi = \int \frac{d P}{F_{1+1}(P)}.
$$
These two relations are actually compatible, since the 1+1 GUP function can be defined as the contraction of the 3+1 one via
$$
F_{1+1} (P)= f(P^2) +P^2 g(P^2) = \frac{f^2(P^2)}{f(P^2) - 2 P^2 f'(P^2)}
$$
implying indeed that 
$$
\int \frac{d P}{F_{1+1}(P)} = \frac{P}{f(P^2)}.
$$
}:
\beq
\label{relfonda}
\Pi_{i}=\frac{P_{i}}{f(\PP^{2})}.
\eeq
Note that this relation must be considered as formal, as long as it has not been shown explicitly that this operator can be properly represented on $\Hilb$; more on that later. With such a relation, we can easily derive two other main equations, namely
\beq
\label{fonda2}
F_{ij} = \frac{\partial P_j}{\partial \Pi_i}, \quad \frac{\partial \Pi_i}{\partial P_j} =  F_{ij}^{-1}
\eeq

Using the generators of translations and rotations introduced above, it is then straightforward to see that any Hamiltonian invariant by translation and rotations (i.e. $[H,\bpi]=0$ and $[H,\LL]=0$), which shall represent the free particle in 3+1, must be of the form $H(\PP^2)$. Here $\LL=(L_1,L_2,L_3)$ and the same for $\bpi$. This is the form we will adopt hereafter. 

On the other hand, time dependent symmetries, corresponding to (Galilean) boosts $G_{i}$ in the case of the free particle, must satisfy
\beq
\label{boostdef}
i \frac{\partial}{\partial t} G_i = [H(\PP^2), G_i].
\eeq
The general solution of Eq. (\ref{boostdef}) is not easy to determine in general. However, when we further restrict to an operator linear in $\XX$, we get the most general solution to be:
\beq
G_i = A_{ij}(\PP) X_j - 2 t A_{ij}(\PP) (f+ g \PP^2) H'(\PP^2) P_{j},
\eeq
where the $A_{ij}$'s are analytic functions. Note that boost generators linear in the position operator match our expectation of the affinity of space, i.e. of the homogeneity of space. We can define implicitly another Hamiltonian, by using Eq. (\ref{relfonda}):
\beq
H_{\Pi}(\bpi^2) \equiv H(\PP^2).
\eeq
Then the boosts can also be written as 
\beq
\label{boostenpi}
G_i = A_{ij} \left( X_j - 2 t  H_{\Pi}' \Pi_{j} \right),
\eeq
where we used the useful formula $\partial \bpi^2/\partial \PP^2 = f(\PP^2)- 2 \PP^2 f'(\PP^2)/f(\PP^2)^3$ that can be derived using Eq. (\ref{relfonda}). Also, we defined $H_{\Pi}' \equiv \partial H_{\Pi}(\bpi^2)/ \partial \bpi^2$.

The boosts are therefore determined only up to an unknown tensor $A_{ij}$. This must not come as a surprise. For instance, the $\bpi$'s are not the only generators that are ``translation like''. In fact, any function $B_{ij}(\PP) P_j$ are also valid generators of symmetries (since they commute with $H$). We usually choose in particular the $\bpi$'s as ``true'' generators of translation because they are the only ones that shift wave-functions $\psi(\mathbf{x}) \to \psi(\mathbf{x+a})$, which is equivalent to requiring the transformation Eq. (\ref{translation}) on averaged values. But this requires both an Hilbert space representation, and some extra physical input. 

Moreover, here we simply cannot choose the standard boosts $G_i = - m X_i + t \Pi_i$, or equivalently the commutators $[G_i,X_j] = i t \delta_{ij}$ and $[G_i,\Pi_j] = - i m \delta_{ij}$, for it is easily seen that Eq. (\ref{boostenpi}) would then impose $H_{\Pi} = \bpi^2/(2m)$. The theory in this case would be trivial, i.e. standard quantum mechanics but written in terms of a complicated momentum variable. A genuinely deformed theory requires deformed boosts. This is why the correspondence principle here cannot be used at all to fix their form.

So, up to now, boosts are naturally defined only up to some unknown function that we can choose freely. What we have above is the class of boost-like symmetries. To select what we actually call boost requires some further input. 

Finally, one may verify, through lengthy but straightforward calculations, that the Jacobi identities between all the $\XX,\PP,\bpi,\LL,\GG,H$, with obvious notations for $\GG$, are identically satisfied, thus not providing any more constraints.

\section{Different equivalent algebraic formulations}

In this section, we want to emphasise that there exists a degeneracy between the functional form of $H$, the choice of a ``physical'' momentum, and the algebra between $\XX$ and $\PP$. Consequently, the boosts are also affected in this change of variable(s). In practice, this means that {\bf at the level of the algebra}, without further reference to the representation of these algebras on $\Hilb$, we have classes of theories that are indistinguishable from each other, i.e. related by a pure redefinition of the relevant operators\footnote{Note, that these mappings relating the various algebras of operators are not algebra homomorphisms, due to their lack of linearity, in general.}.

Indeed, let us start with the previous model, i.e.: 
\beq
H(\PP^2), \,\, [X_i,P_j]= i \left(f \delta_{ij}+ g P_i P_j \right) 
\eeq
with the generators of symmetry:
\begin{align}
&\Pi_i= P_i/f(\PP^2), \, \, L_i = \epsilon_{ijk} X_j \Pi_k,\nonumber  \\
& G_i =  A_{ij}(\PP) X_j - 2 t A_{ij}(\PP) (f+ g \PP^2) H'(\PP^2) P_j 
\end{align}
satisfying a deformed Galilean algebra:
\begin{align}
& [\Pi_i,\Pi_j] =0, \, [\Pi_i,L_j] = \varepsilon_{ijk} \Pi_k, \, [L_i ,L_j] = \varepsilon_{ijk} L_k \nonumber  \\
& [\bpi, H]=0, \, [\LL,H]=0, \, [\GG, H]=- i \dot \GG, \nonumber  \\
& [\Pi_i, G_j]= - i A_{ji}, \, [L_i, G_j]=- i \epsilon_{iab}\left( A_{jb} X_a + 2 t A_{ja} H_{\pi}' \Pi_b\right), \nonumber  \\ 
& [G_i,G_j]= 2  i \left( A_{ia} \frac{\partial A_{jb}}{\partial \bpi^2}- A_{ja} \frac{\partial A_{ib}}{\partial \bpi^2}\right) \Pi_a \left(X_b - 2 t H_{\Pi}' \Pi_b \right) \nonumber
\end{align}
while the non trivial actions on space and momentum operators are:
\begin{align}
& [X_i, G_j]= 2  i \frac{\partial A_{ja}}{\partial \bpi^2} \Pi_i\left(X_a - 2 t H_{\Pi}' \Pi_a\right) \\
& \qquad \qquad \, \, - 2 i t A_{ji} H_{\Pi}' - 4 i t A_{ja} \Pi_i\Pi_a H_{\Pi}''  \\
&[P_i, G_j] = - i A_{jk }F_{ki}, 
\end{align}
Now we define some transformation:
\beq
\label{RedefMomenta}
\tilde{P}_i = u(\PP^2) P_i
\eeq
Then it generates another GUP of the same form (see details in the appendix A):
\begin{align}
&\tilde{H}(\tilde{\PP}^2) = H (\PP^2(\tilde{\PP}^2)) \\
&[X_i, \tilde{P}_j]= i (\tilde{f} \delta_{ij}+ \tilde{g} \tilde{P}_i \tilde{P}_j)
\label{NewComm}
\end{align}
where the relation between $\tilde f$ and $f$ is given implicitly by
\begin{align}
\tilde \PP^2 \equiv u^2(\PP^2) \PP^2 \\
\tilde f(\tilde{\PP}^2 ) = u(\PP^2) f(\PP^2)
\end{align}
and $\tilde g$ has the same relation to $\tilde f$ than $g$ to $f$. The generators also keep their form:
\begin{align}
&\Pi_i= \frac{\tilde{P}_i}{\tilde{f}(\tilde{\PP}^2)}, \, L_i = \epsilon_{ijk} X_j \Pi_k, \\
\tilde{G}_i &=  \tilde{A}_{ij}(\tilde{\PP}) X_j - 2 t \tilde{A}_{ij}(\tilde{\PP}) (\tilde f + \tilde g \tilde \PP^2) \tilde{H}'(\tilde{\PP}^2)\, ,
\label{NewBoosts}
\end{align} 
but where $\tilde A$ needs not be related to $A$ (because it only comes as an integration "constant" in the general solution, linear in $\XX$, of $i \,\partial_t \tilde{\GG} = [\tilde H, \tilde \GG]$. The main point, however, is that we can freely change the Hamiltonian, and it shall simply generates a new function $f$, or vice-versa. Therefore, the GUP algebra is intrinsically linked to a deformed dynamics in the quantum gravity regime. This is most clearly seen, of course, when one focuses on the transformation from $H=\PP^2/2m, \, [\XX,\PP] \neq i \delta$ to a non quadratic Quantum Mechanics $[\XX,\bpi] = i \delta$, but with a Hamiltonian given by $H(\bpi^2) \neq \bpi^2/2m$. This is done in the next subsection.

\subsection{Aquadratic Quantum Mechanics}

One of the most interesting application of the previous results is the following. Choose $u(\PP^{2})=1/f(\PP^{2})$ in the change of momentum variables (\ref{RedefMomenta}). Then the new GUP, Eqs (\ref{NewComm})-(\ref{NewBoosts}) is characterised by $\tilde f =1$ and $\tilde g=0$. Thus $\tilde P_i = \Pi_i$ and $H_\Pi(\Pi^2) = H(\PP^2(\Pi^2))$.

Any given $f$, and any given $H(\PP^2)$ therefore generates a new Hamiltonian in the translation generators. In the very specific case where $H(\PP^2)=\PP^2/2m$ usually encountered in the literature, the previous relations can be inverted explicitly: $\Pi^2 = H_\Pi^{-1} (\PP^2/ 2 m)$, and thus the aquadratic Hamiltonian is related to the GUP function $f$ by:
\beq
\label{mastereq}
f(\PP^2) =  \sqrt{\frac{\PP^2}{H^{-1}_\Pi(\PP^2/2m)}},
\eeq
which is one of the main results of this paper. This way, one generates a non trivial aquadratic Hamiltonian which is universal. This discussion thus recovers, and enlarges, the discussion found in \cite{Das:2008kaa} about universal corrections of quantum gravity via the GUP.

In this case, the boost operators take the form:
\beq
\tilde{G}_i =  A_{ij}(\Pi) X_j - 2 t A_{ij}(\Pi) H_\Pi '(\Pi^2) \Pi_j 
\eeq
and 
\begin{align}
&[X_i,\tilde{G}_j] =  i \partial_{i} (A_{jk}) X^k - 2 i t  \partial_{i}\left( A_{jk} H_{\Pi}'(\Pi^2) \Pi_k \right) \nonumber \\
&[\Pi_i, \tilde{G}_j] = - i A_{ji} \nonumber  \\
&[\tilde{G}_i, \tilde{G}_j] =  i A_{ia} \partial_{a} A_{jb} X_b  + 2 i t A_{ja}  \partial_{a}(A_{ib} H_{\Pi}'(\Pi^2) \Pi_b) - (i \leftrightarrow j).
\end{align} 
Hence, the boosts sector cannot be standard (in the sense that $\GG= - m \XX + t \bpi)$ unless: 
\beq
- 2 A_{jk} (\delta_{ik} H_{\Pi}' +2 H_{\Pi}'' \Pi_i\Pi_k) = \delta_{ij},
\eeq
which clearly imposes $H_{\Pi}''=0$, thus a quadratic Hamiltonian in $\Pi^2$, and a diagonal $A_{ij}$, i.e., standard Quantum Mechanics. In other words, the only Quantum Mechanics that preserves the standard expression of Galilean invariance is, up to a change of variable, standard Quantum Mechanics. On the other hand, one can obtain a modified Quantum Mechanics provided one deforms the algebra of boosts.

\subsection{QM on Deformed Space}
There is however a way to recover exactly standard QM plus usual Galilean symmetry, starting from the GUP. The price to pay however, is to introduce a momentum dependent space coordinates, very much like in the spirit of DSR theories (here we somehow deal with its quantum version; note that this is a loose analogy and by no means a proven link between this form of GUP and DSR).

Once again, one starts from the standard GUP as described above. Also, consider that $H= \PP^2/2m$ as usual in the literature (if not, change $f \to \tilde f$ such that it is the case). Then, define $A_{ij}$ such that:
\beq
A_{ij} \equiv - m F^{-1}_{ij}.
\eeq
Using some formul\ae \,like
\beq
F^{-1}_{ij} = \frac{1}{f} \left( \delta_{ij} - 2 \frac{f'}{f} P_i P_j \right)
\eeq

and defining new coordinates:
\beq
\tilde X_i = F^{-1}_{ij}(\PP) X_j,
\eeq
one gets a standard boost operator:
\beq
G_i = - m \tilde X_i + t P_i .
\eeq
All calculations done, one then obtains an equivalent model with:
\beq
H= \frac{\PP^{2}}{2m},
\eeq
and:
\begin{align}
&[\tilde X_{i},P_{j}] = i \delta_{ij}\\
&[\tilde{X}_{i},G_{j}] = it\delta_{ij}\\
&\left[P_{i},G_{j}\right]=im\delta_{ij}.
\end{align}
Moreover, one can show that
\beq
[G_i, G_j] =0, \quad [\tilde X_i, \tilde X_j]=0
\eeq
This is just standard, Galilean, quantum mechanics on deformed, momentum dependent, space coordinates. Note, however, that this equivalence would be greatly challenged in presence of an interaction potential involving the position operator.

\subsection{Short Summary}
We have shown that, at the algebraic level, there exists a degeneracy between the choice of Hamiltonian, $H$ and the choice of the momentum operator, thus impacting the structure of the boost sector of the theory. The complicated, quadratic GUP with non-trivial boosts and GCR can always be cast in the form of aquadratic Quantum Mechanics with $\Pi$-dependent boosts, standard commutation relations, but aquadratic Hamiltonian. On the other hand, by changing $X$ and not $P$ this time, we can also cast any GUP in the form of pure quadratic Quantum mechanics with exact Galilean symmetries, but on momentum dependent space operators.

The case of a more general Hamiltonian that would describe a particle in some force field $H(X,P)$ is more involved, and we expect, as some sort of quantum generalization of Darboux theorem, that such transformations always exist as well in this case, although their explicit form may be difficult to write down in general.

\section{Hilbert space representations}

The previous study of algebraic properties of the GUP is valid at the formal level of algebra, but might present some complications when the objects considered are represented as linear operators on some Hilbert space, in particular because functional calculus must be used\footnote{Functional calculus basically means making sense of functions of operators.}, and the domains of the newly formed operators must be carefully studied.

In this section, we will concentrate on the Hilbert space representations of the first two cases considered previously: GUP in the standard form (modified algebra with standard Hamiltonian) and aquadratic Quantum Mechanics, leaving the third case, i.e. Galilean Quantum Mechanics on deformed space, for future work.
In the following, we will always assume that operators on the Hilbert space $\mathcal{H}$ are closed (necessary to be able to use functional calculus) and densely defined (necessary to be able to speak about the adjoint of an operator).

\subsection{Standard 3+1 GUP : Hilbert space representation}

This section generalizes Kempf et al. representation in 3+1 dimensions \cite{Kempf:1994su}. This means that we work in the momentum representation, for some function $F_{ij}$ and $H=\PP^2/2m$. Then the position and momentum operators are represented by:
\bey
P_i \psi(\pp) = p_i \psi(\pp)\\
X_i = i F_{ij}(\pp) \frac{\partial}{\partial p_j} \psi(\pp) 
\eey
One may check that $[X_i,X_j] \psi(\pp) = 0$ is satisfied because of the specific form of $F_{ij}$,  $F_{ij} = f \delta_{ij} + g p_i p_j$. Now, we shall assume that the $\PP$'s are self-adjoint. This might seem trivially true because they are multiplication operators, but it holds only if their spectrum is the full space $\mathbb{R}^3$); more on this below. On the other hand, the $X_i$'s are made symmetric\footnote{A symmetric operator $A$ on a Hilbert space $\mathcal{H}$ is a linear operator which is defined on a domain $\mbox{dom} (A)$ which is dense in $\mathcal{H}$ and such that: $\forall(x,y)\in\mbox{ dom }(A)^{2}, \bra{Ax}\ket{y}=\bra{x}\ket{Ay}$.} provided the following scalar product is used:
\beq
\bra{\psi}\ket{\phi} = \int \frac{d^3 p}{\det F} \psi^*(\pp) \phi(\pp)
\eeq
The Hilbert space is thus :
\beq
\mathcal{H}=L^2\left(\mathbb{R}^3, \frac{d^3p}{\det F}\right),
\eeq
i.e. the vector space of square integrable (complex) functions defined on $\mathbb{R}^{3}$ with an unconventional integration measure $\frac{d^3p}{\det F}$, provided of course that the measure is itself well-defined on $\mathbb{R}^3$. This depends on the precise shape of the chosen function $f$, since we have $\det F = f^4/ (f - 2 \pp^2 f')$. For now, we shall thus assume that $1/\det F$ does not vanish nor diverge anywhere except maybe at infinities (see also the concluding remarks). Then $\PP$ are selfadoint on $\R^3$.
 
The previous claim is easily proven by computing first:
\beq
\bra{X_i\psi}\ket{\phi}= \int \frac{d^3 p}{\det F} (- i) F_{ij} \frac{\partial }{\partial p_j }\left(\psi^*(\pp)\right) \phi(\pp) ,
\eeq
where we used the partition of unity:
\beq
\mathbb{I}=\int \frac{d^3 p}{\det F} | \pp >  < \pp |
\eeq
Integrating by parts, one gets:

\beq
\bra{X_i\psi}\ket{\phi} = \bra{\psi}\ket{X_i\phi} +  i \int d^3 p \,  \psi^*(\pp) \phi(\pp) \frac{\partial }{\partial p_j }\left( \frac{F_{ij}}{\det F}   \right). 
\eeq
The last term is zero because
\bey
\frac{\partial}{\partial p_j }\left( \frac{ F_{ij}}{\det F} \right) =\frac{1}{\det F} \left( \partial_j F_{ij} - \frac{F_{ij}\partial_j \det F}{\det F}  \right)
\eey
vanishes. Indeed, using $\det F = exp(tr \ln F)$, we have $\partial_j \det F= \det F \, Tr ( F^{-1} \partial_j F)$, or $\partial_j F_{ij} = F_{ik} \partial_k \ln \det F$.

\subsubsection{Self-adjointness of the $\XX$ operator and existence of a minimal length}
Although they can be made symmetric, as emphasized in \cite{Kempf:1994su}, the $\XX$ operators need not be self-adjoint in general\footnote{Remember that the condition for an operator to be self-adjoint is stronger that the condition to be symmetric: a linear operator $A$ on a Hilbert space $\mathcal{H}$ is self-adjoint if and only if it is symmetric and its domain is equal to the domain of its adjoint.}. Recall that it is necessary to have self-adjoint operators in order to be able to apply the spectral theorem to them. Moreover, as we shall see below, self-adjointness (or lack thereof) is a crucial property with respect to the existence or not of a minimal length in these models.

The self-adjointness of $\XX$ actually strongly depends on the choice of the function $f$, as we now show. We follow closely arguments provided in \cite{Kempf:1993bq}. To this end, we use the von Neumann method of deficiency indices. By definition, the deficiency subspaces for each $k\in\{1,2,3\}$ are \cite{Weyl:1910weyl,vonNeumann:1929vNeu} (see e.g. \cite{Bonneau:1999zq} for an introduction to the use of this result):


\beq
L_k^{\pm}=Ker(X_k^\dagger \mp i)= \{\psi \in D(X_k^\dagger) | \bra{X_k \psi}\ket{\phi} = \pm i \bra{\psi}\ket{\phi} \forall \phi \in \Hilb\} .
\eeq

Here $D$ refers to the domain of the operators considered. The deficiency indices $n_{k, \pm}$ are the dimensions of $L_k^{\pm}$. A symmetric operator is self-adjoint iff its deficiency indices are both zero.
Using the formulas above, we find that
\beq
L_k^{\pm}=\left\{\psi \in D\left(X_{k}^\dagger\right) | \int\frac{d^3 p}{\det F} \left(iF_{kl} \frac{\partial \psi^*(\pp)}{\partial p_l} \mp i \psi^*(\pp)\right) \phi(\pp) =0,\,\forall \phi \in \mathcal{H}\right\} 
\eeq
Deficiency spaces are thus spanned by functions that are normalizable and solutions of: 
\beq
F_{kl} \frac{\partial \psi^*(\pp)}{\partial p_l} = \pm \psi^*(\pp).
\eeq
The only solutions are found using the generator of translations and  Eq. (\ref{fonda2}):
\beq
\psi_{k, \pm}(\pp) = c \, e^{\pm \pi_k(\pp)},
\eeq
for each $k$, and where $c \in\mathbb{C}$.
These states have a norm given by
\beq
|c|^2 \int \frac{d^3 p}{\det F}  e^{\pm 2 \pi_k(\pp)} = |c|^2 \int d^3 \pi e^{\pm 2 \pi_k},
\eeq
after a change of variable $p_i \to \pi_i$ which we assume to be smooth enough. While the range of integration over the $p_i$'s is $\mathbb{R}^3$ because $\PP$ are self-adjoints on $\Hilb$, it needs not be the case for the range of the $\pi_i$'s. 

More precisely, the deficiency indices $n_{k, \pm}$ for each $k\in\{1,2,3\}$ are zero (i.e. the operators $X_{k}$ are self-adjoint) provided these functions are not normalizable, i.e. if and only if the $\pi_k$'s diverge when the $p_k$'s do. In 1+1 dimension, this requires that $\pi$ is in a one to one relationship from $\mathbb{R}$ to $\mathbb{R}$, while in 3+1, this is is equivalent to say that the $\pi^2$, as a function of  $\pp^2$, is in a one to one relationship from $\mathbb{R}^+$ to $\mathbb{R}^+$. In this case,  the $X_k$ operators are essentially self-adjoint. On the contrary, if the $\pi_k(\pp)$'s are bounded functions, the $X_k$'s operator are not self-adjoint\footnote{These are the only two possible cases. Indeed the relation $\pi_k = p_k/f(\pp^2)$ reduces to $\pi_k = p_k/f(p_k^2)$ in the limit $p_k \to \pm \infty$, and thus is antisymmetric at infinities. Thus the range of the functions $\pi_k$ must be either of the form $]-\infty, +\infty[$, in which case deficiency indices are $(0,0)$, or $]-A,A[$ with $A$ a finite number, in which case deficiency indices are $(1,1)$.}.

\subsubsection{Minimal length and the function $f$ in the GCR}

In order to illustrate the previous section, let us go back to the specific case considered in \cite{Kempf:1994su}, i.e. the standard one dimensional GUP
with $[X,P]= i\left(1+ \beta P^2\right)$, we have: 
\beq
\label{KMM}
\Pi = \frac{\arctan \left( \sqrt{\beta} P \right)}{\sqrt{\beta}}
\eeq
as an operator-valued relation, defined by a functional calculus of the self adjoint operator $P$. Therefore, the spectrum of $\Pi$, which is also the range of the function $\pi(p)$, is bounded from above and below. In this case the deficiency indices are $(1,1)$, and thus the $X$ operator is not self-adjoint but admits self-adjoint extensions; see \cite{Kempf:1993bq}. It is pretty clear, both intuitively and formally, that this also implies a minimal length. Indeed, since $[X,\Pi] = i$, we have $\Delta x \Delta \pi \geq 1/2$, but the existence of a maximal $\pi_{\textrm{max}}$ imposes $\Delta \pi \leq \pi_{max}$, and thus a minimal length
\beq
\Delta x \geq \frac{1}{\pi_{\textrm{max}}}
\eeq
This holds in general (and also in 3+1). Indeed, for any normalized state $\psi.\psi =1$, the mean value of $\bpi^2$ reads
\bey
< \bpi^2 >  &=& \int   \frac{d^3 p}{\det F} \psi^*(\pp) \pi^2(\pp^2) \psi(\pp)  \nonumber \\
&\leq&  \pi_{\textrm{max}}^ 2 \int \frac{d^3 p}{\det F} \psi^*(\pp) \psi(\pp) = \pi_{\textrm{max}}^ 2 
\eey
In turn, a bounded value for the translation operator signifies a minimum amount of permissible translations, i.e. a kind of quantization of space. As mentioned in the introduction, this has been discussed in specific contexts in particular in \cite{ali:2009zq,das:2010zf}. Here we have shown that these ideas are very general, and we see that the key element to know whether this holds is the range of the generators of translations as functions of the physical momentum.

On the other hand, (and still in 1+1), a self-adjoint position operator requires $Ran(\pi) = Ran(\int^p dp'/F_{1+1}(p'))= ]-\infty, +\infty[$. In this case, it is always possible to construct physical states (i.e. states that belong to $\Hilb$) such that their $\Delta x \to 0$, see \cite{Kempf:1994su}. These GCRs thus do not imply a minimal length.

In 3+1, the criteria is also about the range of the $\pi_i$'s, and thus actually about the range of the function $\pi^2$ as a function of $\pp^2$. What is relevant there is whether the function $x/f^2(x)$, defined on $\mathbb{R}^+$, diverges or converges at infinity. 

\subsection{Aquadratic Quantum Mechanics : Hilbert space representation}
We now ask ourselves whether the theory obtained by the Hilbert space representation of the Aquadratic Quantum Mechanics has any chance to be equivalent to the previous theory. This is not obvious a priori. The issue is not settled yet if a potential is also present, but holds for the free particle as we now show. Let us give the basics of such a representation.\\
\\
From the Stone-von Newman theorem, the representation of $[X_{k},\Pi_{l}]= i \delta_{kl}$ is unique up to a unitary transformation. We thus basically have standard Quantum Mechanics representation with:
\begin{align}
X_i \psi = x_i \psi  \\
\Pi_{j} \psi = - i \nabla_j \psi
\end{align}
on $\mathcal{H} = L^2(\mathbb{R}^3, d^3x)$ in the position space representation. Thus $\XX$ and $\bpi$ are self-adjoint, and their spectra are $\mathbb{R}^3$. This seems to be in contradiction with the previous section, where we found that the translations can have maximal eigenvalues. 

However, we now introduce a non quadratic Hamiltonian $H_{\Pi}(\bpi^2)$ into the theory, which will further restrict the space of physical states. Since $\bpi^2$ itself is self-adjoint on some relevant domain, the Hamiltonian $H_{\Pi}$ can thus be defined by a functional calculus, depending a bit on its injectivity properties.
The key ingredient here is thus the restriction of the space of physical states for which $H_{\Pi}. \psi$ are both defined and normalizable (if not, we could not write down the Schrödinger equation). 

By definition, we have
\beq
 H_{\Pi} | \psi > = \int d^3 \pi H_{\Pi}(\pi^2) \psi(\pi) | \pi >
\eeq
and thus, in the case where $H_{\Pi}$ has a bounded domain of definition, we have to restrict accordingly the range of the above integral. Therefore the space of physical wave-functions can no longer be defined on the whole set $\pi \in \mathbb{R}^3$, but only $]-\pi_{\textrm{max}},\pi_{\textrm{max}}[^3$ where the aquadratic Hamiltonian is itself defined\footnote{We also have to check that $H_{\Pi}.\psi$ are normalizable. This requires the convergence of
$$
\int d^3 \pi H_{\Pi}^2(\pi^2) \psi^*(\pi) \psi(\pi) < \infty
$$
In the case of a minimal length, or bounded $\pi$, the Hamiltonian may diverge at the boundaries, and thus physical states must vanish fast enough near $\pm \pi_{\textrm{max}}$.}. This implies in turn $\Delta \pi \leq \pi_{\textrm{max}}$. For instance, in the standard one dimensional GUP, one gets, using Eq.~(\ref{KMM}):
\beq
H_{\Pi} = \frac{\tan \left(\sqrt{\beta} \Pi\right)^2}{2 m \beta}.
\eeq
The dynamics can thus only be defined on the domain $]-\frac{\pi}{2 \sqrt{\beta}},\frac{\pi}{2 \sqrt{\beta}}[$, where, for once,  $\pi$ here is the actual number $3.14$ etc. and not the generator of translation.

More generally, the link with the previous section should now be clear. Because we have the following relation between the translation operators and the physical momenta, 
\beq
\bpi^2 = H_{\Pi}^{-1}\left(\frac{\PP^2}{2 m}\right),
\eeq
we see that having the $\pi^2(\pp^2)$ ranging from $]-\pi_{\textrm{max}},\pi_{\textrm{max}}[$ is actually equivalent to having an Aquadratic Hamiltonian $H_{\Pi}$ defined only on the cube $]-\pi_{\textrm{max}},\pi_{\textrm{max}}[^{3}$ or an open subset of that cube, hence resulting in an upper-bound on the generator of translations which, in turn, leads to a minimal length, as seen previously. 

We thus see that here, the minimal length emerges, not from modified commutation relation, but rather from the requirement that the Hamiltonian is a well-defined operator on the set of physical states which then becomes a subset of the original permissible Hilbert space.

If one now wishes to go the other way around, ie. from AQM to the standard GUP, this is sighlty more involved. Indeed, although $\bpi$ are self adjoint on $\R^3$, they cannot be self adjoint anymore on the restricted Hilbert space of the physical states that must have bounded eigenvalues $\pi_i$'s. This is similar to what happens in the quantum mechanics of a particle in a box, where the position operator is not self adjoint anymore but admits self adjoint extensions. Hence, in order to go from AQM to the GUP, one would first need to consider the self adjoint extensions of $\bpi$, and then, and only then\footnote{This is because a functional calculus can only be defined on self adjoint operators.}, define by a functional calculus the physical momenta $\PP$'s via the inversion of Eq.~(\ref{relfonda}).

\section{Conclusion}

In this paper, we have derived an algebraic equivalence between the standard `commutative‘ GUP with a quadratic Hamiltonian and an infinite number of different algebraic structures, and showed explicitely how symmetries transform accordingly.

In particular, we showed how the standard GUP can be recast in the form of a standard Heisenberg algebra, provided one switches to an aquadratic Hamiltonian. In either cases, the (Galilean) boosts must be non-standard, otherwise the theory reduces to the usual Quantum Theory of the free particle, possibly written in non-standard variables. A non-trivial theory thus impose energy (ie. momentum) dependent boosts, implying somehow that different observers see different spacetimes, as in DSR phenomenology. We made this point  very clear when we showed that the theory can be written, at least at the algebraic level, as standard quantum mechanics but in terms of momentum-dependent space 'coordinates' (operators, in fact).

Furthermore, we studied conditions under which such an algebraic equivalence can be maintained once one tries to represent these algebras on a given Hilbert space of physical states. Such a result would amount to showing that all these representations are equivalent up to a unitary transformation. This is a non trivial statement, and therefore we have limited ourselves to whether or not a given algebraic formulation of the GUP displays a minimal length scenario. Indeed, we have provided some necessary and sufficient conditions in order for a standard GUP and the corresponding aquadratic QM to both lead to a minimal length. Specifically, we have shown that the key property that needs to be elucidated in any given model is the relation $\pi(p)$ between the eigenvalues of the generators of translations and the eigenvalues of the physical momentum of particles. This relation actually determines everything else, ie. both the aquadratic Hamiltonian in the AQM framework, or the GCRs in the GUP framework, but also the spectral properties of the $X$ operators, and thus the presence or not of a minimal length and/or of a minimal translation.

Because of this result, there are only four main cases from a mathematical point of view, all of them describing quite different qualitative physics. In order to make this conclusion simple enough, we restrict here to the one dimensional case (but generalization has been shown to be straightforward). In this paper, we have mainly focussed on these first two cases:  
\begin{enumerate} 
\item[1.] $\pi(p)$ is one-to-one from $\R$ to $\R$; or 
\item[2.] $\pi(p)$ is one-to-one from $\R$ to $]-\pi_{\textrm{max}},\pi_{\textrm{max}}[$.
\end{enumerate}
In particular, we have assumed that $P$ is self adjoint on the real line. In the first case, we showed that $\Pi$ and $X$ are self-adjoint on $\R$ and that there is no minimal length in the theory. Also, $H(\pi)$ is well-defined on the whole of $\R$ and does not diverge at any finite value of $\pi$. In the second case, where translations are bounded from above, $X$ turns out not to be self-adjoint: it has deficiency indices $(1,1)$. A minimal length appears, and  $H$ diverges at the boundaries $\pm\pi_{\mbox{max}}$. 

There are however two other cases that would deserve some more work, ie. the cases where the spectrum of the physical momentum is itself bounded $]-p_{\textrm{max}},p_{\textrm{max}}[$. As we discussed above, these cases manifest themselves when discussing the well-posedness of the measure $d^3p/ \det F$ on the Hilbert space. In particular, this happens if the GUP function $f$ is defined only on some finite domain; eg. $f = 1+ x/(1-x^2)$. Following our reasoning so far, we may conjecture the following results:
\begin{enumerate}
\item[3.] $\pi(p)$ is one-to-one from $]-p_{\textrm{max}},p_{\textrm{max}}[$ to $\R$, ie. somehow the opposite of case 2. Then we expect $\Pi$ and $X$ to be self adjoint, but not $P$. There is no minimal length, but a bounded momentum and Hamiltonian. 
\item[4.] Finally, if $\pi(p)$ is one-to-one from $]-p_{\textrm{max}},p_{\textrm{max}}[$ to $]-\pi_{\textrm{max}},\pi_{\textrm{max}}[$,  none of these operators are self-adjoint. The theory exhibits a minimal length, a maximum momentum and energy. 
\end{enumerate}
These statements should however be studied in more details. As a matter of fact, as discussed at the end of the last section, we recall that a functional calculus can only be defined on self-adjoint (or normal) operators. Therefore self-adjoint extensions must be first considered before making sense of Eq. (\ref{relfonda}). This is why we have restricted ourselves to the first two cases, and leave these two last for future works.

Finally, we remark that if $\pi(p)$ is multivalued, then the possiblity of inverting this relation requires to restrict the study to relevant intervals on which we always get back to one of the previous cases. Each interval defines a separate physics.

From the point of view of phenomenology, this means that the De Broglie formul\ae \, must be modified, such that these four cases would naturally describe the following different properties for a wave carrying quanta (bearing in mind, of course, that the GUP presented here is non relativistic):
\begin{enumerate}
\item[-] In case 1, a wave can have $\lambda \to 0$ when its quanta have infinite energy $p \to \infty$;
\item[-] In case 2, a wave is such that $\lambda \to 1$ when its quanta have energy $p \to \infty$;
\item[-] In case 3, a wave can have $\lambda \to 0$ when its quanta have energy $p \to 1 $;
\item[-] In case 4, a wave is such that $\lambda \to 1$ when its quanta have energy $p \to 1$;
\end{enumerate}
in natural units, and up to numerical factors.

Besides the last two cases that we discussed briefly in this conclusion, the rest of the work presented here would also need to be further studied when a potential is present, and/or when space coordinates do not commute anymore. Also, the Galilean Quantum Mechanics with momentum dependent position operators deserves to be studied on its own for its potential link to DSR-like physics.
As mentioned in the introduction, the physical consequences of these quite different behaviours in the UV are to be discussed in a separate publication, with a comprehensive analysis of the particle in a 3D box given any function $f$. Of particular interest here are the questions of discretization of space, finiteness of the Hilbert space, holographic properties, and effective spectral dimension of the theory.

\begin{acknowledgements}
JL is supported by the National Research Foundation (South Africa). JPB thanks N. Lombard for enlightening discussions.
\end{acknowledgements}

\appendix

\section{Derivation of Eqs (\ref{NewComm})-(\ref{NewBoosts})}
We have: 
\begin{eqnarray}
[X_i, \tilde{P}_j ] &=& [X_i , u  P_j ] = i u F_{ij} + 2 i u' F_{ia} P_a P_j \\
 &=& i u f \delta_{ij} + i \left( u g +2  u' (f+ g \PP^2) \right) P_i P_j\\
&=& i \left( \tilde{f} \delta_{ij} + \tilde{g} \tilde{P}_i \tilde{P}_j \right).
\end{eqnarray}
By identification then:
\begin{eqnarray}
\tilde f(\tilde \PP^2) &\equiv& u(\PP^2) f(\PP^2)\\
  \tilde g &\equiv& \frac{g}{u} + 2 (f+g \PP^2) \frac{u'}{u^2}\\
   &=& 2 f \frac{f' u + f u'}{u^2(f -2 \PP^2 f')}.
\end{eqnarray}
And the relation between $\tilde f$ and $\tilde g$ must be checked. For this, note that we have 
\beq
\tilde \PP^2 = u^2(\PP^2) \PP^2
\eeq
Thus 
\beq
\frac{d \tilde \PP^2}{d \PP^2} = u^2 + 2 \PP^2 u u' 
\eeq
Now we should have $\tilde g = 2 \tilde f \tilde f'/(\tilde f - 2 \tilde \PP^2 \tilde f')$ where $'$ here refers to $d/d\tilde \PP^2$. Compute
\beq
\frac{d \tilde f}{d \tilde \PP^2} = \frac{d (u f)}{d \PP^2}\frac{d  \PP^2}{d \tilde \PP^2} = \frac{(u' f + u f')}{u^2 + 2 \PP^2 u u'}.
\eeq
Simplify:
\beq
\frac{2 \tilde f \tilde f'}{\tilde f - 2 \tilde \PP^2 \tilde f'} = 2 u f  \frac{(u' f + u f')}{u^2 + 2 \PP^2 u u'} \frac{1}{u f - 2 u^2 \PP^2  \frac{(u' f + u f')}{u^2 + 2 \PP^2 u u'} }
\eeq
to
\beq
= 2 f \frac{f' u + f u'}{u^2(f -2 \PP^2 f')}= \tilde g. \quad  \Box
\eeq
Finally, given the unchanged functional form of the GCR, and $\tilde H$, it is necessary that the boosts conserve their form, and the same apply to the computation of $[\XX, \tilde \GG], [\tilde \PP, \tilde \GG]$, etc.

\bibliographystyle{spmpsci}
\bibliography{biblio}


\end{document}